
\magnification=\magstep1
\hfuzz=4truept
\nopagenumbers
\baselineskip=18truept
\font\tif=cmr10 scaled \magstep3

\rightline{PUPT-1530}
\vfil
\centerline{\tif Dimensional continuation without perturbation theory }
\vfil
\centerline{{\rm Vipul
Periwal}\footnote{${}^\dagger$}{vipul@puhep1.princeton.edu}}
\bigskip
\centerline{Department of Physics}\centerline{Princeton University
}\centerline{Princeton, New Jersey 08544-0708}
\vfil
\par\noindent A formula is proposed for continuing physical correlation
functions
to non-integer numbers of dimensions, expressing them as infinite weighted
sums over the same correlation functions in arbitrary integer dimensions.
The formula is motivated by studying
the strong coupling expansion, but the end result makes no reference to
any perturbation theory.  It is shown that the formula leads to
the correct dimension dependence in weak coupling perturbation theory
at one loop.
\medskip
\vfil\eject

\def\be{\beta}
\def\part{\partial}

A question of interest, given the importance of dimensional
regularization[1] in obtaining physical predictions from gauge theories,
is the following:  What is the meaning of dimensional regularization
beyond the Feynman diagrams of weak coupling perturbation theory?
It seems difficult to find a non-perturbative definition of
dimensional regularization, at least while preserving translation
invariance.

I show in this paper that there is a simple dimensional
continuation  which, while motivated by strong  coupling perturbation
theory, gives a formula that is not perturbative in character.  It is,
of course, well-known that strong coupling expansions, or weak coupling
perturbative
expansions, can all be continued in dimension without difficulty.  The
novelty here is that the formula presented does not need perturbation
theory of any sort for its statement.

The strong
coupling origins of this formula are classical---recall that
in the work of Fisher and Gaunt[2], it was shown how to expand
strong coupling series for large $d.$  This required a grouping of
strong coupling graphs in a geometric manner that is rather unnatural
from the point of view of continuum field theory.  Nevertheless, this
grouping of strong coupling graphs, recalled below, is such that the
contribution from each separate group of graphs can be extracted from
the sum of all the graphs, provided one knows the sum of all the graphs in
every dimension smaller than or equal to the integer dimension of interest.
It is then a simple manner to write a dimensional continuation formula
that involves an infinite summation over terms, with each term a finite
sum over quantities at integer dimensions.
Due to the infinite sum, the
formula has a `transcendental' character, probably unavoidable for
a formula valid for arbitrary dimensions.

While the formula is motivated by strong coupling perturbation theory, I
show that it gives the correct answer by performing an explicit computation
in weak coupling field theory.  This calculation gives a different insight into
the manner in which the analytical continuation is working.

I shall first explain the grouping of strong coupling graphs suggested by
Fisher and Gaunt[2].  I shall then give the formula, which is almost
obvious from the strong coupling expansion.  Finally, I show that the
formula gives the correct continuation to non-integer dimensions
for a weak coupling one-loop calculation of a
$\phi^3$ field theory correlation function.

Consider an Ising model on a $d$-dimensional hypercubic lattice.  For a
connected
spin-spin correlation function, {\it e.g.} $d^{-1}\sum_{j=1}^d
\langle S(0) S(n\hat e_j)\rangle,$
($\hat e_j$ are the basis vectors for the lattice)
the strong coupling, or high temperature, expansion can be organized as
a sum over paths that go from 0 to the point $n\hat e_j$ on the lattice.
For any such path, there is a hyperplane
spanned by the vectors corresponding to the steps in the path, $d_{min}.$
It follows that the correlation function can be written as
$$d^{-1}\sum_{j=1}^d\langle S(0) S(n\hat e_j)\rangle=\sum_{D=1}^{d} C^d_D
\ F_{D}^{irred},$$
where $F_{D}^{irred}$
is defined to be the
contribution from paths that span a fixed hyperplane of dimension $d_{min}=D,$
and the combinatorial coefficient
$$C^d_D\equiv \ {\Gamma(d+1)\over\Gamma(D+1)\Gamma(d-D+1)} $$
counts the number of $D$-dimensional
hyperplanes in $d$ dimensions.  As it stands, this is not particularly
useful, since we have no obvious way of computing the quantity
$F_{D}^{irred}.$  However, it is evident
from the vanishing of $C^d_D$ if $d<D,$ that we can replace the upper limit
in the sum by infinity.  So, if we could compute $F_{D}^{irred}$ for
integer $D,$ our analytic continuation would be
$$d^{-1}\sum_{j=1}^d\langle S(0) S(n\hat e_j)\rangle=\sum_{D=1}^{\infty} C^d_D
\ F_{D}^{irred}.$$

Now, given a sum of the form
$$ a_d =\sum_{D=1}^{d} C^d_D b_D,$$
it is easy to check that
$$b_d =\sum_{D=1}^{d} (-)^{d-D} C^d_D a_D.$$
It follows that our desired analytical continuation is just
$$d^{-1}\sum_{j=1}^d\langle S(0) S(n\hat e_j)\rangle=\sum_{D=1}^{\infty} C^d_D
\sum_{k=1}^{D} (-)^{D-k} C^D_k
\left[k^{-1}\sum_{l=1}^k\langle S(0) S(n\hat e_l) \rangle\right].$$
We can now state the general formula for dimensional continuation of
suitable physical quantities:
$$ F_d =\sum_{D=1}^{\infty} C^d_D (-)^D\
\sum_{k=1}^{D} (-)^{k}\ C^D_k\ F_k;$$
in words, for a physical quantity, $F,$ that can be described in a
dimension-independent way, and which has the same engineering dimensions
in all dimensions, given the value of $F$ at integer dimensions, the value at
an arbitrary complex dimension can be computed.  When $d$ is an integer, it is
easy to check that we recover the identity $F_d=F_d.$
\def\ee{{\hbox {e}}}

\def\al{\alpha}
\def\be{\beta}
To develop an intuition for how this analytical continuation is working,
I shall
now compute the dimensional continuation of the two-point function at
one-loop in $\phi^3$ theory.  This is a computation in weak coupling
perturbation theory, and therefore completely independent of the
strong coupling motivations for the formula.  The calculation must be done
with a regularization which works in any number of dimensions.   This requires
rapid decay of propagators with increasing momenta, or working with a
lattice regularization.  Working on
a hypercubic lattice in $D$ dimensions,
$$G(p) = \left(m^2+{2\over a^2}\sum_{1}^{D}(1- \cos ap_j)\right)^{-1}$$
is the lattice Green function.
The one-loop correction to the two-point function is
$$ \Gamma^{(2)}_{1-loop,D}(p) =
- {1\over 2} g^2 a^{D-6}\int_{-\pi/a}^{\pi/a} {d^Dq\over(2\pi)^D}
G(q)G(p+q),$$
which can be written as
$$- {1\over 2} g^2 a^{D-6}\int_{-\pi/a}^{\pi/a} {d^Dq\over(2\pi)^D}
\int_0^\infty d\al \int_0^\infty
d\be \exp\left[-\al G^{-1}(q) -\be G^{-1}(p+q)\right].$$
Suppose that $p=(p_1,\dots),$ then we have
$$\Gamma^{(2)}_{1-loop,D}(p) = - {1\over 2} g^2 a^{-6}\int d\al d\be
\ee^{-(\al+\be)(m^2+2D/a^2)}
I_0\left({2(\al+\be)\over a^2}\right)^{D-1} \widehat I(\al,\be,p_1),$$
where $I_0$ is a Bessel function, and
$$\widehat I(\al,\be,p_1) \equiv \int^{\pi/a}_{-\pi/a} {dq_1\over 2\pi}
\exp\left[{2(\al+\be)\over a^2}\cos(aq_1) + {4\be\over a^2}\sin(a(p_1/2+q_1))
\sin(ap_1/2)\right].$$
This quantity is now analytic in $D,$ so we don't need the analytic
continuation formula given---however, the question we are attempting to
answer is whether this same analytic continuation will arise from the
dimensional continuation formula given above.

Plugging in, we obtain, interchanging integration and summation,
$$\eqalign{ \Gamma^{(2)}_{1-loop,d}(p) =
 - {1\over 2} g^2a^{-6} &\int d\al d\be\
\widehat I(\al,\be,p_1)I_0\left({2(\al+\be)\over a^2}\right)^{-1}
\cr &\times \sum_{D=1}^{\infty} C^d_D\sum_{k=1}^{D}
(-)^{D-k} C^D_k  \ee^{-(\al+\be)(m^2+2k/a^2)}
I_0\left({2(\al+\be)\over a^2}\right)^{k}\cr
= - {1\over 2} g^2a^{-6} &\int d\al d\be\ \ee^{-(\al+\be)m^2}
\widehat I(\al,\be,p_1)I_0\left({2(\al+\be)\over a^2}\right)^{-1}
\cr &\times \sum_{D=1}^{\infty} C^d_D (-)^D
\left(\left[1-I_0\left({2(\al+\be)\over a^2}\right)\ee^{-2(\al+\be)/a^2}
\right]^D -1\right)\cr
= - {1\over 2} g^2 a^{-6}&\int d\al d\be\
\ee^{-(\al+\be)(m^2+2d/a^2)}
I_0\left({2(\al+\be)\over a^2}\right)^{d-1} \widehat I(\al,\be,p_1) .\cr}$$
As advertized, the dimensional continuation formula gives back the
expected answer, even in this weak coupling calculation.

The point of this simple example is, however, that it shows that
the seemingly complicated dimensional continuation formula is essentially
just a formula for defining arbitrary powers of a number, $x,$ as a series in
its integer powers, by expanding it in a Taylor series about $1,$ in
integer powers of $(1-x),$ and then expanding each integer power of
$(1-x)$ as a polynomial in $x.$  Contrast this with the strong coupling
derivation.

How unique is this dimensional continuation?  Let us investigate different
ways of continuing $x$ to arbitrary complex powers, {\it e.g. }let $\lambda$
be a complex number and consider the sum
$$\eqalign{f_\lambda(d) &\equiv \sum_{D=1}^{\infty} C^d_D\sum_{k=1}^{D}
(-)^{D-k} C^D_k x^k \exp\left(\lambda(d-k)\right)\cr
&=\ee^{\lambda d}\sum_{D=1}^{\infty} C^d_D\sum_{k=1}^{D}
(-)^{D-k} C^D_k \left(x\ee^{-\lambda}\right)^k\cr
&=\ee^{\lambda d} \left(x\ee^{-\lambda}\right)^d = x^d.\cr}$$
Thus, for functions that grow at most geometrically with dimension, we
get back the same function for any choice of $\lambda,$ consistent with
$|1-x\ee^{-\lambda}|<1.$  (Recall that
according to Carlson's theorem[3], if two functions agree on integers,
and separately satisfy the bound $|f_i(d)| < \exp(c|d|),$ for $\Re d>0,$
with $c<\pi,$
then they define the same analytic function.)

One may wonder if the double summation in the continuation formula is
really necessary.  It would be much more pleasant if one could work with
a series of the form
$$ f(d) = \sum_{D=1}^{\infty} f(D) c(d,D),$$
with some coefficients $c(d,D).$  Indeed, if one looks at the finite
sub-sum occurring in each term of the infinite sum, we see that using the
vanishing of the combinatorial coefficients $C^j_k$ for $k>j,$ we can
write this as an infinite sum as well:
$$ F_d =\sum_{D=1}^{\infty} C^d_D (-)^D
\sum_{k=1}^{\infty} (-)^{k} C^D_k F_k,$$
Now, if we ignore the fact
that we are rearranging terms in a double summation with terms that
have alternating signs, we arrive at a formula of the desired type, with
the unfortunate property that $c(d,D)=\delta_{d,D},$ in other words, the
dimensional continuation is not analytic.  It would therefore appear that
the double sum is necessary, and that the finiteness of the nested summation is
necessary for the analyticity of the formula.

A possible application of this formula is to gauge theory.  The only
gauge-invariant regulators known are dimensional regularization, and
lattices.  It would be of some interest to study a dimensionally
continued lattice formula in the limit as the cutoff goes to zero, or
as the dimension approaches 4.  A more interesting possibility would be
to interpret the infinite summation in terms of embedding $d$-dimensional
Euclidean space in an infinite-dimensional space, with an appropriate
$d$-dependent measure, such that integration with respect to this
measure would lead to the dimensional continuation formula given in this
paper.  While I have no evidence that this can be achieved, it would
be one approach to making concrete non-perturbative sense of dimensional
continuation for field theories.

\bigskip
Acknowledgements:  I am grateful to D. Gross and R. Myers for
helpful conversations.  This work was supported in part by NSF Grant
No. PHY90-21984.
\bigskip
\centerline{References}
\bigskip
\item{1.} J. Ashmore, {\it Lett. Nuovo Cim.} {\bf 4} (1972) 289; G.'t Hooft
and M. Veltman, {\it Nucl. Phys.} {\bf B44} (1972) 189; C.G. Bollini and J.J.
Giambiaggi, {\it Phys. Lett.} {\bf 40B} (1972) 566.
\item{2.} M.E. Fisher and D.S. Gaunt, {\it Phys. Rev.} {\bf 133} (1964) 225
\item{3.} E.C. Titchmarsh, {\sl The Theory of Functions}, Oxford University
Press (1939)
\end